\documentclass[twocolumn]{article}
\usepackage{graphicx,natbib}

\title{\bf Magnetic fields of active galactic nuclei and quasars with polarized broad H$\alpha$ lines}
\author{N.A. Silant'ev$^1$, Yu.N. Gnedin$^{1,2}$\thanks{E-mail:gnedin@gao.spb.ru}, S.D. Buliga$^1$, \\M.Yu. Piotrovich$^1$ and T.M. Natsvlishvili$^1$\\
$^{1}$Central Astronomical Observatory at Pulkovo of Russian Academy of Sciences, \\Pulkovskoye chaussee 65, Saint-Petersburg, 196140, Russia\\
$^{2}$St.-Petersburg State Polytechnical University, Polytechnicheskaya 29, \\Saint-Petersburg, 195251, Russia}
\begin{document}

\maketitle

\begin{abstract}
We present estimates of magnetic field in a number of AGNs from the Spectropolarimetric atlas of Smith, Young \& Robinson (2002) from the observed degrees of linear polarization and the positional angles of spectral lines (H$\alpha$) (broad line regions of AGNs) and nearby continuum. The observed degree of polarization is lower than the Milne value in a non-magnetized atmosphere. We hypothesize that the polarized radiation escapes from optically thick magnetized accretion discs and is weakened by the Faraday rotation effect. The Faraday rotation depolarization effect is able to explain both the value of the polarization and the position angle. We estimate the required magnetic field in the broad line region by using simple asymptotic analytical formulas for Milne's problem in magnetized atmosphere, which take into account the last scattering of radiation before escaping from the accretion disc. The polarization of a broad spectral line escaping from disc is described by the same mechanism. The characteristic features of polarization of a broad line is the minimum of the degree of polarization in the center of the line and continuous rotation of the position angle from one wing to another. These effects can be explained by existence of clouds in the left (keplerian velocity is directed to an observer) and the right (keplerian velocity is directed from an observer) parts of the orbit in a rotating keplerian magnetized accretion disc. The base of explanation is existence of azimuthal magnetic field in the orbit. The existence of normal component of magnetic field (usually weak) makes the picture of polarization asymmetric. The existence of clouds in left and right parts of the orbit with different emissions also give the contribution in asymmetry effect. Assuming a power-law dependence of the magnetic field inside the disc, we obtain the estimate of the magnetic field strength at first stable orbit near the central supermassive black hole (SMBH) for a number of AGNs from the mentioned Spectropolarimetric atlas.
\end{abstract}


\section{Introduction}

Smith et al. (2002) presented  optical spectropolarimetric atlas of 36 nuclei of Seyfert 1 Galaxies. The data were obtained with the William Herschel and the Anglo--Australian Telescopes from 1996 to 1999. It is well-known that spectropolarimetry is an important tool in studies of active galactic nuclei (AGN). The spectropolarimetric data provide the detailed view into the inner regions of active galactic nuclei, including an accretion disc and accretion flows around a supermassive black hole (SMBH), thus allowing one to probe the structure and kinematics of the polarizing material around the accreting SMBH.

Smith et al. (2002) objects exhibit a variety of characteristics with the average degree of polarization ranging from $0.2$ to $5$ percent. They show many variations both in the degree $p_{line}(\lambda)$ and position angle $\chi_{line}(\lambda)$ of polarization across the broad H$\alpha$ emission line. The characteristic feature of $p_{line}(\lambda)$ is the minimum in the line centre, which is usually less than the polarization degree $p_c(\lambda)$ in nearby continuum. The second feature is the monotonic increase of positional angle from one line wing to the other. Note also that there exists little difference in the mean polarization degrees and position angles of nearby continuum and H$\alpha$ line for nine of Seyfert galaxies (Mrk 6, Akn 120, Mrk 896, Mrk 926, NGC 4051, NGC 6814, NGC 7603, UGC 3478, ESO 012-G21). For 22 measurements out of 45, the mean positional angles $\chi_{line}$ and $\chi_c$ are practically the same. It should be emphasized that the position angles of polarized continuum and H$\alpha$ line coincide more frequently than their polarization degrees. The mean value of the polarization degree in continuum over all sources is 0.68\%, and the same value for H$\alpha$ lines is 0.66\%. A position angle is most sensitive to the geometry of emitting region. For that reason, it is most probable that both emitting regions are located near one another, i.e. they located in an accretion disc and their scale sizes are similar: $R_{BLR}\approx R_{\lambda}$, where $R_{\lambda}$ corresponds to the scale size of an accretion disc for the continuum radiation with given wavelength $\lambda$.

It is generally accepted that AGNs are powered by the release of gravitational energy from gas accreted onto supermassive black holes (SMBH). The well-known anti-correlation between the radius of the broad-line region (BLR) and the velocity width of broad emission lines for AGNs supports the idea that the BLR gas is virialized and that its velocity is dominated by the gravity of the SMBH (Peterson \& Wandel 2000; Onken \& Peterson 2002).

Most of the recent results lead to the conclusion that BLR presents a flattened rotating system. Many authors (Vestergaard et al. 2000; Nikolajuk et al. 2006; Sulentic et al. 2006; La Mura et al. 2009; Bon et al. 2009; Punsly \& Zhang 2010) pointed out that considerable flattening and a predominantly planar orientation are likely to be the intrinsic property of the BLR structure. This conclusion allows us to consider BLR as an outer part of geometrically thin accretion disc that is optically thick with respect to the electron Thomson and the Rayleigh scattering processes.

Seyfert galaxies were traditionally divided into two classes according to the presence or absence of broad optical lines. Antonucci \& Miller (1985) explained this phenomenon by obscuration by a dusty torus with different orientation with respect to an observer. The orientation-based unification model has become quite popular, but it has also been confronted by more specialized observations (see, for example, Zhang \& Wang 2006; Wang \& Zhang 2007); in particular, there is evidence for the existence of a special subclass of Seyfert 2 lacking hidden broad-line regions (Zhang \& Wang 2006). Thus, the paradigm of unification scheme for all Seyfert galaxies remains a matter of debate (Miller \& Goodrich 1990; Tran 2001, 2003).

The basic feature of Smith et al. (2002, 2004, 2005) models is that the polarization plane for most of Seyfert galaxies is parallel to the direction of the radio jet. Simultaneously, these models postulate that the radio jet direction is perpendicular to the accretion plane. The latter assumption is questionable. In reality, the radio jets frequently have significant bends near the radio core. The angle of the bend depends on the ratio of radial and toroidal magnetic fields in the accretion disc. Besides, the direction of the jet appears to change with time (see, for example, Britzen et al. 2009; Rastorgueva et al. 2011). Therefore, the coincidence of the direction of the radio jet and the polarization plane does not mean that the electrical vector of the polarized radiation ${\bf E}$ is perpendicular to the accretion disc. Of course, one can introduce the angle between the radio jet and the electric field ${\bf E}$ as an additional characteristic of AGNs. But, strictly speaking, this angle is not necessarily related to the real inclination of ${\bf E}$ with respect to the accretion disc. As a last resort, this angle may be considered in a probabilistic sense.

The existence of many cases when the position angle of radiation has intermediate value between parallel or perpendicular to the direction of the radio jet also demonstrates that real direction of ${\bf E}$ does not correlate with the jet direction. As a result, we conclude that the models describing the polarization behavior in AGNs should not assume that the radio jets are perpendicular to accretion discs; instead, the explicit dependence on the inclination angle $i$ of the accretion disc needs to be taken into account.

It is commonly accepted (see, for example, Blaes, 2003) that the accretion discs are magnetized. The existence of radio jets is usually associated with strong magnetic fields in centers of AGNs and quasars. Numerous theoretical models demonstrate the power-law dependence of the magnetic field distribution in an accretion disc. Usually accretion discs are considered as geometrically thin slabs with Thomson optically depth $\tau \gg 1$. The scattering-induced linear polarization can be as high as $\sim 12\%$ for edge-on viewing (Chandrasekhar 1960). However, in the real situation of a magnetized accretion disc, the degree of polarization $p$ will be reduced due to Faraday rotation of the radiation polarization plane while a free photon travels between the consequent scatterings. Recall that the angle of Faraday rotation $\Psi$ at the Thomson optical length $\tau$ is equal to

\begin{equation}
 \Psi=\frac{1}{2}\delta\tau\cos\Theta,\,\,\,\,
 \delta=0.8\lambda^2 B,
 \label{eq1}
\end{equation}

\noindent where the wavelength of radiation $\lambda$ is measured in microns and the magnetic field $B$ is measured in Gauss. The angle $\Theta$ is the angle between the direction of light propagation ${\bf n}$ and the direction of ${\bf B}$.

The decrease of the polarization degree due to Faraday rotation occurs as a result of summation of chaotic angles of rotations in the multiple scattering process. This process has been considered in many papers (for example, Silant'ev 1994; Agol \& Blaes 1996; Gnedin \& Silant'ev 1997). Clearly, the value $\Psi\sim 1$ at the mean free path $\tau\sim 1$ can decrease considerably the standard Chandrasekhar's polarization degree. Besides, the dependence of $\Psi$ on the wavelength and magnetic field gives rise to characteristic dependencies of the polarization degree $p$ and the position angle $\chi$ of radiation on $\lambda$, which allows us to estimate the strength and direction of the magnetic field in the scattering region.

Below we develop a new model for the formation of polarization in AGNs, which does not use the assumption of the position angle of observed radiation being correlated with the direction of the radio jet. We hypothesize that the observed polarization is due to intrinsic polarization of radiation outgoing from the magnetized optically thick accretion disc (the Milne problem in magnetized atmosphere). In our model, the characteristic features of polarization mentioned above are explained by the topology of the magnetic field in the accretion disc, when the Faraday rotation of the polarization plane is taken into account. Primarily, we suppose that the whole radiating surface of the magnetized accretion disc is observed, i.e. that the inclination angle $i$ is such that the obscuring torus (if it really exists) does not intersect the line of sight. We also consider the case when we observe only a part of total surface of accretion disc, i.e. we take into account the obscuring torus. It appears that our model and the usual pure geometrical model of Smith et al. (2002, 2004, 2005), which takes into account single scattering of BLR-photons in nearby clouds, are two competing explanations for the polarization properties of the accretion discs.

All actually observed polarization degrees $p_c$ are much smaller than the value in the Milne problem in non-magnetized atmosphere at the same inclination angle. Recall that the Milne problem deals with the radiative transfer in an optically thick atmosphere, where the sources of thermal radiation are located far from the surface, at depth with $\tau\gg 1$. In optically thick accretion discs the main source of thermal radiation is found at the midplane of the disc, and the outgoing radiation is described by the solution of the Milne problem. The mean value of $p_c$ in the atmosphere with pure electron scattering is equal to 3.1\%, and the maximum value is 11.7\%. The outgoing radiation in case of an absorbing atmosphere has a much greater polarization, because the intensity peaks near the surface. In this case, most of polarization arises analogously to the process of a single scattering of a radiation beam near the surface. For the Milne problem in spectral lines, the value $p_{line}$ is less than in continuum (see, for example, Ivanov et al. 1997).

For the accretion disc models, the main challenge is to determine the scale length of the disc - i.e.\ the radius where the disc temperature matches the rest frame wavelength of the monitoring band. A semi-empirical method for measuring the disc scale length has been developed (Kochanek et al. 2006; Morgan et al. 2006, 2008; Poindexter, Morgan \& Kochanek 2008). These authors used microlensing variability, observed for gravitationally lensed quasars, to find the accretion disc scale length for a given observed (or rest-frame) wavelength. Clearly, such a scaling has to be consistent with the most popular accretion disc model of Shakura \& Sunyaev (1973). As a result, Poindexter et al. (2008) presented the following relation for the scale length of a standard geometrically thin accretion disc:

\begin{equation}
 R_{\lambda}=10^{9.987}
 \left(\frac{\lambda_{rest}}{\mu m}\right)^{4/3}
 \left(\frac{M_{BH}}{M_{\odot}}\right)^{2/3}
 \left(\frac{L_{bol}}{\varepsilon L_{Edd}}\right)^{1/3} cm.
 \label{eq2}
 \end{equation}

The wavelength dependence, $R_{\lambda}\sim\lambda_{rest}^{4/3}$, corresponds to the typical (for Shakura--Sunyaev disc model) effective temperature: $T_e=T_{in}(R/R_{in})^{-3/4}$, where $R_{in}$ is the inner radius of an accretion disc and $T_{in}$ is the temperature corresponding to that radius. Here $L_{Edd}=1.3\cdot 10^{38}(M_{BH}/M_{\odot})$ erg\,s$^{-1}$ is the Eddington luminosity, $M_{BH}$ is the black hole mass, $\varepsilon$ is the rest-mass radiation conversion efficiency, and $L_{bol}$ is the bolometric luminosity.

Numerous papers provided measurements of BLR sizes for AGNs (see, for example, Peterson et al. 1994, 2004; Wu et al. 2004; Bentz et al. 2009; Shen \& Loeb 2010; Greene et al. 2010). Kaspi et al. (2007) have compiled the observational data for Seyfert galaxies and nearby quasars with black hole masses estimated with the reverberation mapping technique.

Most recently Shen \& Loeb (2010) have suggested an empirical analytic formula for $R_{BLR}$ that is very useful for various estimates and applications:

\begin{equation}
 R_{BLR}=2.1\cdot 10^{17}M_8^{1/2}\left(\frac{L_{bol}}{L_{Edd}}\right)^{1/2}.
\label{eq3}
\end{equation}

\noindent Here $M_8=M_{BH}/10^8M_{\odot}$. We will use this formula in our further calculations.

Below we estimate the magnetic field strength in BLR of AGN and QSO from the data from the spectropolarimetric atlas presented by Smith et al. (2002). The $\lambda$ -- dependence of the observed polarization degree and position angle in H$\alpha$ line is very complicated. It appears to be produced by large-scale chaotic motions in the accretion disc.

\section{Basic equations}

To estimate the degree of polarization $p$ and the position angle $\chi$ of radiation escaping from the magnetized atmosphere we use the standard radiative transfer equations for Stokes parameters $I, Q$ and $U$ (see, for example, Silant'ev 1994; Dolginov, Gnedin \& Silant'ev 1995; Silant'ev 2002, 2005). This system of equations has a fairly complicated form. Numerical solutions have so far been obtained only for the case when magnetic field ${\bf B}$ is parallel to the normal ${\bf N}$ to an atmosphere (see Silant'ev 1994; Agol and Blaes 1996; Shternin et al. 2003).

For our purpose, however, it is sufficient to use a simple asymptotic theory, which can be presented in an analytical form for an arbitrary direction of the magnetic field in the atmosphere (Silant'ev 2002, 2005; Silant'ev et al. 2009). In this approximation, the intensity of the radiation $I(z,\mu)$ obeys a usual transfer equation with the Rayleigh phase function, and the system of equations for parameters $Q$ and $U$ can be presented in the following form:

\[
\mu\frac{d}{dz}(-Q+iU)=
\]

\begin{equation}
= - \alpha[1+C+i(1-q)\delta\cos\Theta](-Q+iU) +B_Q(z,\mu),
\label{eq4}
\end{equation}

\noindent where $B_Q(z,\mu)$ describes the source function for parameter $Q$ due to the contribution of intensity scattering in non-magnetized atmosphere (in this case $B_U\equiv 0$), $\mu={\bf nN}$ is the cosine of the angle between the direction of light propagation ${\bf n}$ and the normal ${\bf N}$ to the atmosphere, $\alpha$ is the total extinction factor due to Thomson scattering and pure absorption on dust particles, the value $q = \sigma_a / (\sigma_a + \sigma_s)$ (Silant'ev et al., 2009) is the degree of absorption, $C$ describes the additional extinction of polarized radiation due to the fluctuating component ${\bf B'}$ of the magnetic field in the atmosphere (see below). Eq.(4) is valid in the limit of large Faraday rotation parameter $\delta \ge 1$.

A solution of Eq.(4) results in the following expression for parameters $Q({\bf n},{\bf B})$ and $U({\bf n},{\bf B})$ for the radiation escaping from the magnetized atmosphere:

\[
-Q({\bf n},{\bf B})+iU({\bf n},{\bf B})=
\]

\begin{equation}
=-\int^{\infty}_0 \frac{d\tau}{\mu}B_Q(\tau,\mu) \exp{\left(-[1+C+i(1-q)\delta\cos\Theta]\frac{\tau}{\mu}\right)}
\label{eq5}
\end{equation}

\noindent At $\delta\ge 1$, the first non-zero term of integrating by parts of Eq.(5) gives  rise to the asymptotic expression which has an analytical form and can be used for an arbitrary direction of the magnetic field. For example, for the case $B_Q(0,\mu)\neq 0$ we have:

\[
Q({\bf n},{\bf B})\simeq \frac{B_Q(0,\mu)(1+C)}{(1+C)^2+(1-q)^2\delta^2\cos^2\Theta},
\]
\begin{equation}
U({\bf n},{\bf B})\simeq\frac{B_Q(0,\mu)(1-q)\delta\cos\Theta}{(1+C)^2+(1-q)^2\delta^2\cos^2\Theta}.
\label{eq6}
\end{equation}

For the Milne problem in absorbing atmosphere, a more sophisticated theory (see Silant'ev 1994, 2002) gives rise to the following expressions:

\[
Q({\bf n},{\bf B})\simeq \frac{I(0,\mu)\,p^{(1)}(\mu)(1-s\mu)}{(1+C-s\mu)^2+(1-q)^2\delta^2\cos^2\Theta},
\]
\begin{equation}
\frac{U({\bf n},{\bf B})}{Q({\bf n},{\bf B})}\simeq \frac{(1-q)\delta\cos\Theta}{1+C-s\mu}.
\label{eq7}
\end{equation}

\noindent Here $p^{(1)}(\mu)$ is the polarization degree of outgoing radiation, which takes into account only the last scattering before escaping the atmosphere. The value $p^{(1)}(\mu)$ gives the main contribution to $p(\mu)$ - exact polarization degree of outgoing radiation for a non-magnetized atmosphere. For this reason, below we use the value $p(\mu)$ instead of $p^{(1)}(\mu)$, which for $q=0$ is presented in Chandrasekhar (1960), and for the absorbing atmosphere in Silant'ev (1980). The value $s$ is the root of the characteristic equation, tabulated by Silant'ev (1980). If the degree of the true absorption is small ($q \ll 1$), the parameter $s = \sqrt{3 q}$.

First, let us consider the case when the whole surface of a radiating accretion disc is observed. In this case, the Stokes polarization parameters of continuum radiation $Q_c(\varphi)$ and $U_c(\varphi)$ must be averaged over all azimuthal angles $\varphi$, characterizing the position of a radiating surface element on a circular orbit in the accretion disc ($-\pi\le \varphi \le \pi$). The normal ${\bf N}$ and the direction to an observer ${\bf n}$ are the same for all parts of the accretion disc surface. Therefore, the reference frame of the accretion disc is common to all parameters $Q_c(\varphi)$ and $U_c(\varphi)$ of radiation escaping from the disc. In this case, the averaging procedure consists of integrating these parameters over the azimuthal angle $\varphi$. Note, that $\cos\Theta$ depends on $\varphi$ and the integral over $\varphi$ can be taken analytically only over the interval $(-\pi,\pi$). As a result, the observed values for the degree of polarization and the position angle can be derived analytically. Silant'ev et al. (2009) presented the detailed description of the behavior of these quantities for continuum radiation. The degree of linear polarization of continuum $p_c$ and the position angle $\chi_c$ for an accretion disc can be expressed in the analytical form:

\begin{equation}
 p_c({\bf B},\mu)=\frac{p_c(\mu) (1-s\mu)}{\left[g_c^4+2g_c^2(a^2+b^2)+
(a^2-b^2)^2\right]^{1/4}},
\label{eq8}
\end{equation}

\[
 \tan2\chi_c=\frac{U_c}{Q_c}=
\]

\begin{equation}
 =\frac{2ag_c}{\left(p_c(\mu) (1-s\mu)/p_c({\bf B},\mu)\right)^2+(g_c^2+b^2-a^2)}.
\label{eq9}
\end{equation}

\noindent Here $\mu=\cos i$, where $i$ is the inclination angle (angle between the light propagation direction ${\bf n}$ and the normal ${\bf N}$ to the accretion disc). The degree of polarization $p_c(\mu)$ corresponds to a non-magnetized accretion disc. The value of $p_c(\mu)$ for the continuum radiation presents the classical solution of Milne problem (Chandrasekhar 1960) with $p(0)= 11.7\% $. The value of the position angle $\chi_c=0$ corresponds to oscillations of the wave electric vector perpendicular to the plane (${\bf nN}$).

For spectral lines in isotropic medium the value $p_{line}(\mu)$ depends on the specific quantum numbers of the transitions (see, for example, Chandrasekhar 1960) and on the shape of the line. For a dipole type transition and the Doppler line shape, $p_{line}(\mu)$ in the atmosphere with pure electron scattering has the same functional form as Chandrasekhar value $p_c(\mu)$, with maximum value of 9.44\% instead of 11.71\% (see Ivanov et al. 1997). In isotropic medium we have to average an atom over all orientations. For such medium the transfer equation coincides with the usual system for Rayleigh scattering with an additional term, which describes unpolarized radiation. Such average naturally occurs due to usual thermal motions. Considering various quantum numbers for H$\alpha$ line, we see that this additional term is larger than the Rayleigh scattering term (see Chandrasekhar 1960), and the maximum degree of polarization becomes $\sim 3$\% instead of 9.44\%. Therefore, the observed polarization can be explained if the atmosphere also has pure absorption of H$\alpha$  line (the existence of dust). In absorbing atmosphere the polarization degree $p_{line}$ can be larger than that in the non-absorbing atmosphere.

Broad H$\alpha$ lines presented in atlas Smith et al. (2002) have very high widths, laying in the interval 50 -- 200 angstroms. In such situations the total line (sum of 5 sub levels) can be described by one absorption coefficient with the Doppler shape (see, for example, Varshalovich, Ivanchik \& Babkovskaya 2006; Lekht et al. 2008). Our technique takes into account the Faraday rotation during propagation of polarized radiation after the last scattering before escaping from the atmosphere. The region of broad-line  emission is too far from the center of an accretion disc and has low magnetic fields. For this reason, one does not need to take into account the known Zeeman effect.

The parameter

\begin{equation}
 g_c=1+C-s\mu,
\label{eq10}
\end{equation}

\noindent where the negative term $(-s\mu)$ arises in the Milne problem in absorbing atmosphere (see Chandrasekhar 1960). For small degree of true absorption $q=\sigma_{absorb}/(\sigma_{scattering}+\sigma_{absorb})\ll 1$ the factor $s\simeq\sqrt{3q}$. The regions of the line emission are different from those of continuum radiation. Usually one assumes that the parameter $g_c\simeq 1$ in most of areas of the accretion disc. To explain polarization of the line emission, we have to consider that this emission escapes from optically thick clouds with its own dimensionless parameters $g_{line}$.

The dimensionless parameters $a$ and $b$ describe the Faraday depolarization of radiation:

\begin{equation}
 a=0.8\lambda ^2 \mu B_z,\,\,\,\,
 b=0.8\lambda^2\sqrt{1-\mu^2}B_\bot,
\label{eq11}
\end{equation}

\noindent where $B_z\equiv B_{\parallel}$ is the component of the magnetic field directed perpendicular to the accretion disc surface, and $B_{\bot}=\sqrt{B_{\rho}^2+B_{\varphi}^2}$ is the magnetic field in the accretion disc plane. The component $B_{\bot}$ is perpendicular to $B_{\parallel}$. Due to axial symmetry, the inclination angle $\varphi_*$ ($B_{\varphi}/B_{\rho}=\tan{\varphi_*}$) is constant along a circular orbit. The value $0.8\lambda^2B$ is numerically equal to the Faraday rotation angle at the Thomson optical depth of $\tau=2$, if the polarized radiation propagates along the magnetic field direction. Here and in what follows, we take magnetic field in Gauss and wavelengths in microns.

The dimensionless parameter $C$ describes the real situation in a turbulent magnetized plasma and characterizes a new effect -- additional extinction of the polarized radiation (parameters $Q$ and $U$) due to incoherence of the Faraday rotation in small-scale turbulent eddies (see Silant'ev 2005):

\begin{equation}
 C=0.64\tau\lambda^4\langle\left(B'\right)^2\rangle\frac{f_B}{3},
 \label{eq12}
 \end{equation}

\noindent where $\tau$ is the Thomson optical depth of a turbulent eddy ($\tau\ll1$), $\langle(B')^2\rangle$ is the mean value of fluctuations of the magnetic field, and $f_B\approx1$ is a parameter describing the integrated correlation of the $B'$ values at two-closely spaced points in the accretion disc.

It should be noted that the diffusion of radiation in the inner regions of the accretion disc also produces depolarization due to multiple scatterings. Presence of magnetic field and, therefore, the Faraday rotation effect, only increases the depolarization process. As a result, the polarized radiation emitted by a plane-parallel atmosphere at a specific inclination angle is considerably lower as compared to the classical Chandrasekhar--Sobolev value (see, for example, Gnedin \& Silant'ev 1997). But the main feature of Faraday depolarization is the explicit wavelength dependence for both the polarization degree and the position angle.

It is interesting to note that at $a=b$ the polarization degree $p_c({\bf B},\mu)$ takes a maximum value (the term $(a^2-b^2)^2$ is zero in expression (8)). This effect takes place due to the opposite Faraday rotations from magnetic fields $B_{\parallel}$ and $B_{\bot}$ in some places along an orbit.

All formulas presented above show that polarimetric observations allow us to derive the magnetic field strength and its topology in the BLR region, where the polarized radiation escapes the accretion disc. Using various models connecting magnetic field $B_{H}$ at the black hole horizon with the magnetic field $B_{ms}$ at the first stable orbit $R_{ms}$ nearest to the centre of the system, and then using the power law dependence of the magnetic field from $R_{ms}$ up to $R_{BLR}$, we can estimate the magnetic field strength and the parameters that control it, such as the spin of the black hole $a_{*}$, the conversion efficiency of kinetic into radiative energy $\varepsilon$, and the magnetization parameter $k=P_{magn}/P_{gas}$ ($P_{magn}$ and $P_{gas}$ are magnetic pressure and gas pressure, in accreting plasma, respectively).

Frequently one uses simple formulas for $p_c$ and $\chi_c$, corresponding to particular cases of pure normal ${\bf B}_{\parallel}$ and of pure perpendicular ${\bf B}_{\bot}$:

\begin{equation}
p_c(B_{\parallel},\mu)=\frac{p_c(\mu)(1-s\mu)}{\sqrt{g_c^2+a^2}},\,\,\,\,
\tan{2\chi_c}=\frac{a}{g_c},
\label{eq13}
\end{equation}

\begin{equation}
p_c(B_{\bot},\mu)=\frac{p_c(\mu)(1-s\mu)}{\sqrt{g_c^2+b^2}},\,\,\,\,
\chi_c\equiv 0.
\label{eq14}
\end{equation}

\noindent In the latter case $\chi_c\equiv 0$ due to the axial symmetry of the problem.

Now let us consider the case when the radiating gas along a particular orbit in the accretion disc is partly obscured by some dust cloud (an obscuring torus). Clearly, for pure normal magnetic field ${\bf B\parallel N}$, the unobscured part of the orbit has the same polarization degree $p_c(B_{\parallel})$ and the position angle $\chi_c(B_{\parallel})$ as in the case of completely unobscured orbit (see Eq. (13)). If the magnetic field is toroidal ${\bf B}_{\varphi}$, i.e. is tangent to the orbit, we can derive the following analytical expressions:

\[
 p_c(B_{\varphi})= \frac{p_c(\mu)(1-s\mu)}{\sqrt{g^2_c+b^2_{\varphi}}}f_Q(\varphi_0),\,\,\,\chi_c\equiv 0,
\]

\begin{equation}
 f_Q(\varphi_0)=\frac{1}{\varphi_0}\arctan{\left(\frac{\sqrt{g^2_c+b^2_{\varphi}}}{g_c}\,\tan\varphi_0\right)}.
 \label{eq15}
\end{equation}

\noindent Here the parameter $b_{\varphi}=0.8\lambda^2B_{\varphi}\sqrt{1-\mu^2}$, the angle $\varphi_0$ describes the unobscured part of the orbit (we see the orbit within azimuthal angles $-\varphi_0 \le\varphi \le \varphi_0$). At $\varphi_0=\pi$ (complete orbit), Eq.(15) coincides with Eq.(14). Eq.(15) for $f_Q(\varphi_0)$ is valid for $\varphi_0 \le \pi/2$. For $\pi/2 \le \varphi_0\le \pi$ one can use the relation $f_Q(\varphi_0)=[\pi f_Q(\pi/2)-(\pi-\varphi_0) f_Q(\pi-\varphi_0)]/\varphi_0$. A more detailed derivation of these formulas is given below, in subsection 2.2 (for a spectral line case).

\begin{figure}
 \includegraphics[width=84mm]{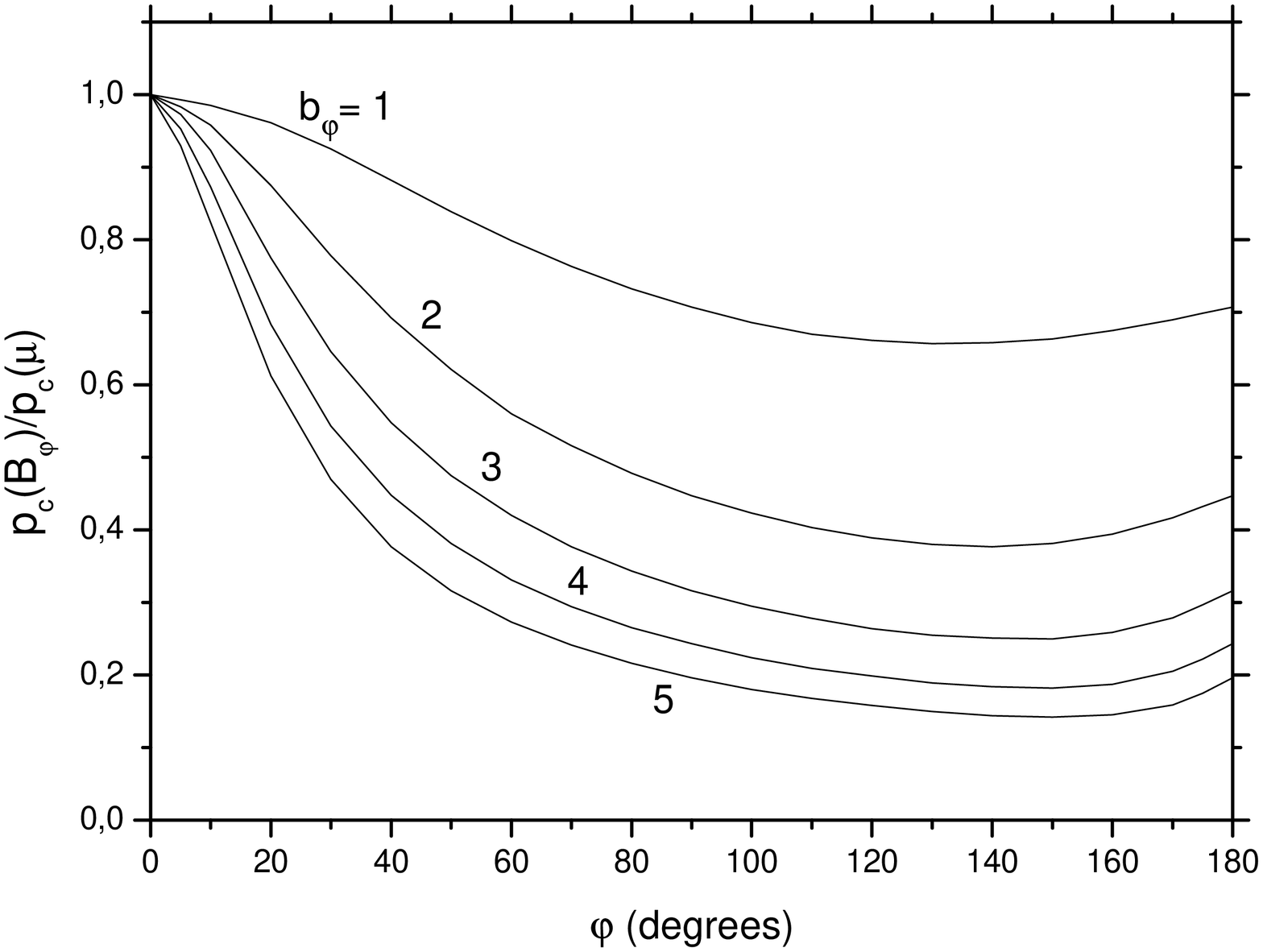}
 \caption{Dependence of $p_c(B_{\varphi})/p_c(\mu)(1-s\mu)$ on $\varphi_0$.}
\end{figure}

Dependence of $p_c(B_{\varphi})/p_c(\mu)(1-s\mu)$ on $\varphi_0$ is presented in Fig.1 for values $g_c=1$, $b_{\varphi}=1,2,3,4,5$, and $s=0$. It is interesting that for $\varphi_0=\pi/2$  (half of the full orbit is observed) the polarization degree coincides with the result (14) for the fully unobscured orbit ($f_Q(\pi/2)=f_Q(\pi)=1$).

\subsection{The case of a spectral emission line}

In the catalog of Smith et al. (2002) the polarimetric data both in the continuum and in the H$\alpha$ emission line are presented. In our model of a rotating accretion disc (with the Keplerian velocity $u_k$) one part (the right side) of the disc ($\varphi =0 \div \pi$) corresponds to motion from an observer and the second part (the left side)  moves towards the observer ($\varphi=\pi \div 2\pi$). According to the Doppler formula, wavelengths of radiation from the first part are greater than the central value $\lambda_0$, and from the second part are smaller than $\lambda_0$. The value $\lambda_0=(1+z)\lambda_{rest}$, where $z$ is redshift parameter of the system and $\lambda_{rest}=0.6563\mu$m is the rest frame wavelength of the H$\alpha$ line.

Here we restrict ourselves to a specific case of a spectral line with the Doppler shape. The line is described
by following normalized shape function:

\begin{equation}
\phi(\lambda)=\frac{1}{\sqrt{\pi}\Delta\lambda_T}\exp{\left[-\left(\frac{\lambda-\lambda_0}
{\Delta\lambda_T}\right)^2\right]}.
\label{eq16}
\end{equation}

\noindent As in the previous case of continuum radiation, we assume that the X-axis is perpendicular to  plane $({\bf nN})$. The Keplerian velocity $u_k$ corresponds to the radius $R_{BLR}$: $u_k=\sqrt{G M_{BH}/R_{BLR}}$, where $G$ is gravitation constant. The usual Doppler line width $\Delta\lambda_T=(u_{turb}/c)\lambda_0$ is mainly due to chaotic turbulent velocities. The displacement of the line centre for radiation emitted from the part of the disc with the azimuthal angle $\varphi$ is $(u_k/c)\lambda_0\sqrt{1-\mu^2}\sin\varphi$. Thus, the normalized shape function of radiation emitted from $\varphi$-part of the orbit has the form:

\[
 \phi(\lambda,\varphi)=
\]
\begin{equation}
=\frac{1}{\sqrt{\pi}\Delta\lambda_T}\exp{\left[-\left(\frac{\lambda-\lambda_0
-\frac{u_k}{c}\lambda_0\sqrt{1-\mu^2}\sin\varphi}{\Delta\lambda_T}\right)^2\right]}.
\label{eq17}
\end{equation}

Observed radiation flux $F_{\lambda}$ from a surface element $dS$ of the ring with the radius $R_{BLR}$ is proportional to $d \varphi$. The flux from the total radiating circular orbit can be obtained by integration over all azimuthal angles $\varphi$. If we suppose that all sources are distributed uniformly along the orbit, this flux is described by the following expression:

\begin{equation}
 F(\lambda)=dS\mu I(\mu)\frac{1}{2\pi}\int^{\pi}_{-\pi}d\varphi\, \phi(\lambda, \varphi).
 \label{eq18}
\end{equation}

\noindent Observed fluxes of linearly polarized radiation differ from the continuum radiation case by an additional factor $\phi(\lambda,\varphi)$:

\[
 F_{Q}(\lambda)=
\]
\begin{equation}
 =dS\mu I(\mu)p_{line}(\mu)(1-s\mu)\frac{1}{2\pi}\int^{\pi}_{-\pi}d\varphi
 \frac{\phi(\lambda,\varphi)g_{line}}{g^2_{line}+\delta^2\cos^2\Theta},
 \label{eq19}
\end{equation}

\[
 F_{U}(\lambda)=
\]
\begin{equation}
 =dS\mu I(\mu)p_{line}(\mu)(1-s\mu)\frac{1}{2\pi}\int^{\pi}_{-\pi}d\varphi
 \frac{\phi(\lambda,\varphi) \delta\cos\Theta}{g^2_{line}+\delta^2\cos^2\Theta}.
 \label{eq20}
\end{equation}

\noindent Here $\Theta$ is the angle between the magnetic field ${\bf B}$ and the light propagation direction ${\bf n}$, $I(\mu)$ is total intensity of the spectral line escaping from the surface $dS$. The azimuthal angle $\varphi=0$ corresponds to a surface element $dS$ perpendicular to the plane $({\bf nN})$. Faraday depolarization term $\delta\cos\Theta$ has the form:

\[
 \delta\cos\Theta=0.8\lambda^2{\bf Bn}=a+b\cos(\varphi+\varphi_*)=
\]
\begin{equation}
=a+b_{\rho}\cos\varphi -b_{\varphi}\sin\varphi,
\label{eq21}
\end{equation}
where the parameters $b_{\rho}$ and $b_{\varphi}$ are:

\[
b_{\rho}=0.8\lambda^2\sqrt{1-\mu^2}B_{\rho}\equiv b\cos\varphi_*,
\]
\begin{equation}
b_{\varphi}=0.8\lambda^2\sqrt{1-\mu^2}B_{\varphi}\equiv b\sin\varphi_*.
\label{eq22}
\end{equation}

\noindent Here angle $\varphi_*$ is the angle between ${\bf B}_{\bot}$ and the radius-vector ${\bf \rho}$, lying in the disc plane. The sign minus before $b_{\varphi}\sin{\varphi}$ corresponds to the right-hand screw rotation of the accretion disc with the frozen magnetic field ${\bf B}_{\varphi}$ directed along the rotation velocity. If the rotation of the disc is opposite, we have to change $b_{\varphi}$ to -$b_{\varphi}$.

If we take the factor $\phi(\lambda,\varphi)=1$ and $g_{line}\to g_c$, all the formulas will describe the case of the continuum radiation. In this case the integrals over $\varphi$ can be evaluated analytically (for the combination $-F_{Q}+iF_{U}$ the $\varphi$-integral can be evaluated by the complex residue method, and we obtain Eqs.(8) and (9)). Note, that these formulas are approximate, they take into account only the last scattering before the escape from the atmosphere. This is a rather satisfactory approximation (see Silant'ev 2002). It describes the main contribution to the polarization. The main merit of these analytical formulas is that they describe the polarization for any direction of the magnetic field. For our purpose this approach is sufficient.

We note that $\varphi$-integration in Eqs.(18) -- (20) gives rise to rather low polarization effects. For this reason they hardly can be used for describing the polarization data presented in the atlas (Smith et al. 2002). Below we present two models that are more acceptable for explaining the data.

\subsection{The model of H$\alpha$ line polarization with parameters $p$ and $\chi$, averaged over the right and left parts of an orbit}

It is clear from Eqs.(18), (19) and (20) that the right parts of circular orbits mostly contribute to gaussian shape lines at wavelengths $\lambda > \lambda_0$, and the left parts mostly contribute to $\lambda < \lambda_0$. Qualitatively, we can consider that these contributions are equivalent to sum of two gaussian shaped polarized lines. We assume that the effective polarizations and position angles of these lines correspond to mean values from the right side ($p_{right}, \chi_{right}$) and the left side ($p_{left}, \chi_{left}$) of the orbit. These values follow from Eqs.(19) and (20) if we take there the factor $\phi(\lambda,\varphi)=1$.

Unlike the situation described by Eqs.(19) and (20), in this model we assume that a part of the accretion disc is invisible due to obscuring torus. The right part corresponds to integration over $\varphi =0 \div \varphi_0$, and the left part corresponds to integration in the interval $\varphi=0 \div -\varphi_0$. Here the angle $\varphi_0$ characterizes the boundary azimuthal angle for the visible part of the BLR orbit. The mean values $\langle Q\rangle$ and $\langle U\rangle$ for visible right part are described by the integrals:

\[
\langle Q_{right}({\bf n},{\bf B})\rangle=
\]

\[
=I_{line}p_{line}(\mu)(1-s\mu)\frac{1} {\varphi_0}\int^{\varphi_0}_0d\varphi\,\frac{g_{line}} {g^2_{line}+\delta^2\cos^2\Theta},
\]

\[
\langle U_{right}({\bf n},{\bf B})\rangle=
\]

\begin{equation}
=I_{line}p_{line}(\mu)(1-s\mu)\frac{1}{\varphi_0} \int^{\varphi_0}_0d\varphi\,\frac{\delta\cos\Theta} {g^2_{line}+\delta^2\cos^2\Theta},
\label{eq23}
\end{equation}

\noindent where $\delta\cos\Theta$ is given in Eqs. (21) and (22). The corresponding mean values for the left part of the BLR orbit are given by integrals in the interval $(0,-\varphi_0)$.

These integrals cannot be evaluated analytically in a general case. We present below the cases ($a\neq 0, b_{\rho}=0, b_{\varphi}=0$) and ($a=0, b_{\rho}=0, b_{\varphi} \neq 0$). The first case corresponds to the magnetic field $B_{\parallel}$ parallel to normal ${\bf N}$. Clearly, in this case the polarization degree and the position angle are the same in the right and left parts of the orbit, and can be obtained
from Eqs.(8) and (9) (see Eq.(13)).

The second case corresponds to a toroidal magnetic field ${\bf B}_{\bot}$, laying in the plane of the accretion disc and tangential to the radiating circular orbit. In this case the Faraday rotations are opposite in the right and the left parts of the orbit. This gives the same value for the polarization degree $p_{right}=p_{left}$ and the opposite values for the position angles $\chi_{right}=-\chi_{left}$. We present below the results for the right part of the orbit:

\[
\langle Q_{right}\rangle = \frac{I_{right}p_{line}(\mu)(1-s\mu)}{\sqrt{g^2_{line}+b^2_{\varphi}}}\,f_Q(\varphi_0),
\]
\begin{equation}
f_Q(\varphi_0) = \frac{1}{\varphi_0} \arctan\left({\frac{\sqrt{g^2_{line}+b^2_{\varphi}}}{g_{line}}\tan{\varphi_0}}\right),
\label{eq24}
\end{equation}
\[
\langle U_{right}\rangle=\frac{I_{right}p_{line}(\mu)(1-s\mu)}{\sqrt{g^2_{line}
+b^2_{\varphi}}}f_U(\varphi_0).
\]
\[
f_U(\varphi_0)=
\]
\begin{equation}
-\frac{1}{2\varphi_0}\ln\left |\frac{\sqrt{g^2_{line}+b^2_{\varphi}}+b_{\varphi}}
{\sqrt{g^2_{line}+b^2_{\varphi}}-b_{\varphi}}\cdot \frac{\sqrt{g^2_{line}+b^2_{\varphi}}-b_{\varphi}\cos\varphi_0}
{\sqrt{g^2_{line}+b^2_{\varphi}}+b_{\varphi}\cos\varphi_0}\right|.
\label{eq25}
\end{equation}

\noindent The expression for $f_Q$ is valid only for $\varphi_0\le \pi/2$, and the formula for $f_U$ is valid for total interval $0\le \varphi_0 \le \pi$. The integrands in Eqs. (23) are symmetric relative to the angle $\varphi_0=\pi/2$. This gives equalities $f_Q(\pi/2)=f_Q(\pi)=1$ and $f_U(\pi/2)=f_U(\pi)$. Due to the aforementioned symmetry, we can calculate values $f_Q$ and $f_U$ for $\pi/2\le \varphi_0\le \pi$ from the values for interval $0\le \varphi_0\le \pi/2$: $f_{Q,U}(\varphi_0)=[f_{Q,U}(\pi)\pi -f_{Q,U}(\pi-\varphi_0)(\pi-\varphi_0)]/\varphi_0$.

It is interesting to note that $f_Q(\varphi_0)$ monotonically grows from $f_Q(\pi/2)=1$ to $f_Q=\sqrt{g^2_{line}+b^2_{\varphi}}/g_{line}\ge 1$ as $\varphi_0\to 0$.

The mean polarization degree $p_{right}(B_{\varphi})$ and the position angle $\chi_{right}(B_{\varphi})$ can be derived from the following expressions:

\[
p_{right}(B_{\varphi})=\frac{p_{line}(\mu)(1-s\mu)}{\sqrt{g^2_{line}+b^2_{\varphi}}}\sqrt{f^2_Q +f^2_U},
\]
\begin{equation}
\tan{2\chi_{right}}=\frac{f_U}{f_Q}.
\label{eq26}
\end{equation}

\begin{figure}
 \includegraphics[width=84mm]{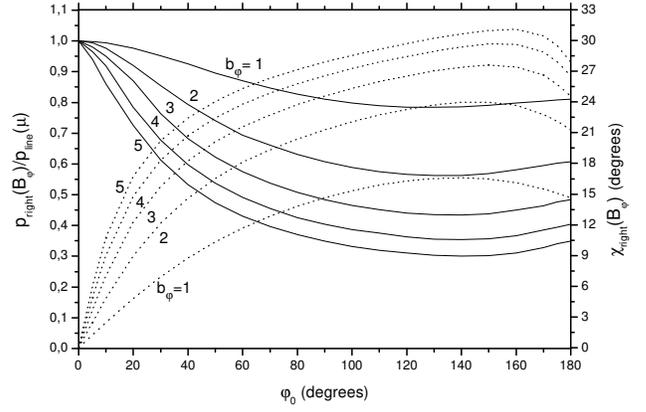}
 \caption{The values $p_{right}(B_{\varphi})/p(\mu)(1-s\mu)$ and $|\chi_{right}|$ (dotted lines) at $b_{\varphi}=1,2,3,4,5$ and $g_{line}=1, s=0$.}
\end{figure}

\begin{figure*}
 \includegraphics[width=170mm]{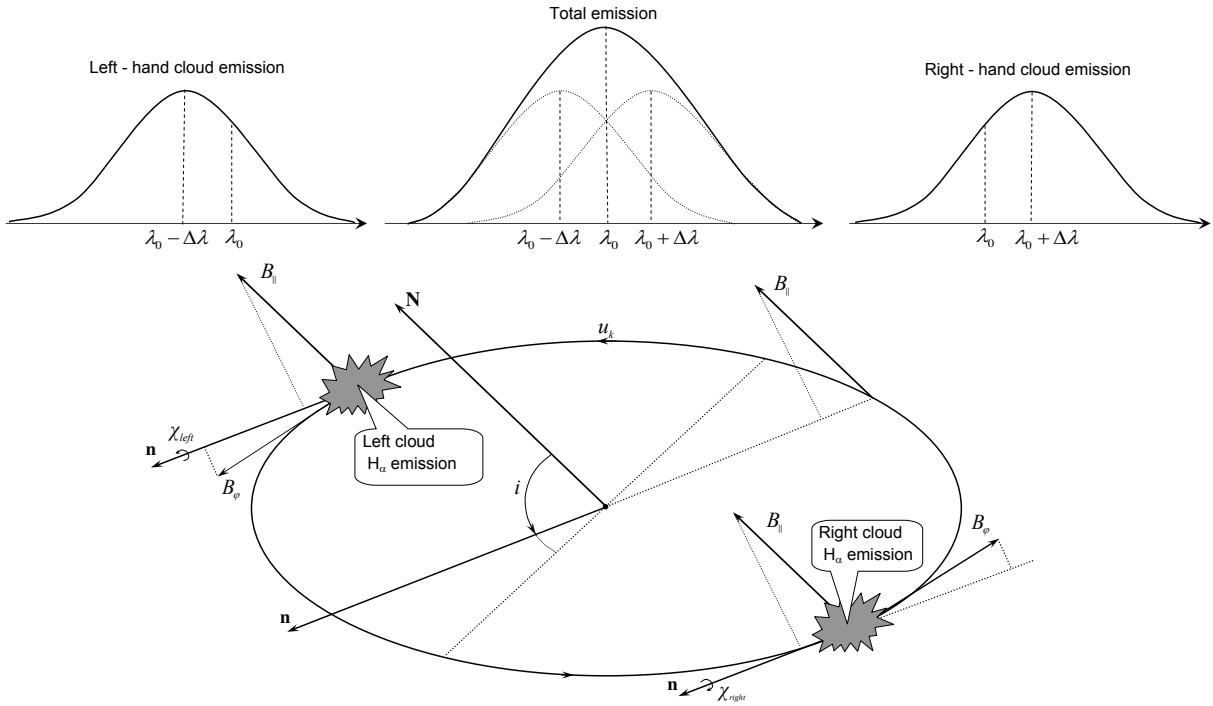}
 \caption{The schematic illustration of the model of two emitting clouds.}
\end{figure*}

\noindent In Fig.2 we present the values $p_{right}(B_{\varphi})/p(\mu)(1-s\mu)$ and $|\chi_{right}|$ at $b_{\varphi}=1,2,3,4,5$ and $g_{line}=1, s=0$. Comparison of $p_{right}(B_{\varphi})$ with $p_c(B_{\varphi})$ in Fig.1 shows that $p_{right}> p_c$. This is evident, because $p_c$ corresponds to the sum of radiation from the right and the left parts of the observed areas (in this sum $U=0$), and $p_{right}$ corresponds to the half of this area, where parameter $U\neq 0$. It means that in the wings of the emitting lines the polarization is greater than in nearby continuum.

Let us consider the behavior of the polarization degree $p_{line}(B_{\varphi},\lambda)$ and the position angle $\chi_{line}(B_{\varphi},\lambda)$ inside the broad spectral line in more detail, using our simple model of two equal gaussian lines with the equal right and left displacements from the central wavelength $\lambda_0$. We label the intensities of these lines as $I_{right}$ and $I_{left}$. According to Eqs.(24) and (25), we present the total observed Stokes parameters $I, Q$ and $U$ in the form:

\begin{equation}
I(\mu)=I_{right}(\lambda)+I_{left}(\lambda),
\label{eq27}
\end{equation}

\begin{equation}
Q(B_{\varphi},\lambda) = (I_{right}(\lambda)+I_{left}(\lambda))\frac{p_{line}(\mu)(1-s\mu)}
{\sqrt{g^2_{line}+b_{\varphi}^2}}\,f_Q,
\label{eq28}
\end{equation}

\begin{equation}
U(B_{\varphi},\lambda) = (I_{right}(\lambda)-I_{left}(\lambda))\frac{p_{line}(\mu)(1-s\mu)}
{\sqrt{g^2_{line}+b_{\varphi}^2}}\,f_U.
\label{eq29}
\end{equation}

\noindent Recall that due to the displacement of the centers of the right and left lines with the gaussian shape the intensities $I_{right}(\lambda)$ and $I_{left}(\lambda)$ are different at a particular considered wavelength inside the full line. Only at the central wavelength $\lambda_0$ these intensities are equal due to the axial symmetry of our model.

Using Eqs.(27), (28) and (29), we obtain the following expressions for the total observed polarization degree $p(B_{\varphi},\lambda)$ and the position angle $\chi(B_{\varphi},\lambda)$:

\[
 p(B_{\varphi},\lambda)=\frac{p_{line}(\mu) (1-s\mu) f_Q}{\sqrt{g^2_{line}+b_{\varphi}^2(\mu)}}\times
\]
\begin{equation}
\times \sqrt{1+\frac{(I_{right}-I_{left})^2}{(I_{right}+I_{left})^2}\left(\frac{f_U}{f_Q}\right)^2},
\label{eq30}
\end{equation}

\begin{equation}
\tan{2\chi(B_{\varphi},\lambda)} = \frac{I_{right}-I_{left}}{I_{right}+I_{left}} \cdot \frac{f_U}{f_Q}.
\label{eq31}
\end{equation}

\noindent Taking in Eqs.(30) and (31) $I_{left}=0$, we revert to Eq.(26) for the right part of the orbit.

The total rotation of the position angle inside the line width is equal to the difference of $\Delta\chi\equiv\chi_{right}-\chi_{left}$, where $\chi _{right}$ corresponds to the right-hand wing of line with $I_{right}\gg I_{left}$. The $\chi_{left}$ corresponds to the left-hand wing with $I_{left}\gg I_{right}$. As a result, we have:

\begin{equation}
|\Delta\chi| =\arctan{\left|\frac{f_U}{f_Q}\right|}.
\label{eq32}
\end{equation}

\noindent Note that the difference $\Delta\chi$ does not depend on the choice of the observer's reference frame. A non-zero value of $\Delta\chi$ is observed in many objects presented in the catalog of Smith et al. (2002) and is due to two reasons -- the presence of the magnetic field $B_{\varphi}$ and the Keplerian rotation of the magnetized accretion disc. Fig.2 shows that $\Delta \chi$ depends on the parameter $b_{\varphi}$ and the angle $\varphi_0$. Thus, for $b_{\varphi} = 5$ and $\varphi_0 \simeq (140^{\circ} - 160^{\circ})$ the value $|\Delta \chi| \simeq 60^{\circ}$.

Now let us discuss shortly the polarization degree $p$ inside the broad line. First of all, we notice that in the centre of the line $\lambda=\lambda_0 $ the polarization is less than in the wings (c.f.\ Eq.(30) with $I_{right}=I_{left}$). The polarization grows with the departure from $\lambda_0 $. This behavior of $p_{line}(\lambda)$ is observed in many objects from the catalog of Smith et al.(2002). According to Eq.(30), the ratio of $p_{wing}(B_{\varphi},\lambda)$ to $p_{center}(B_{\varphi},\lambda)$ becomes

\begin{equation}
\frac{p_{wing}(B_{\varphi},\lambda)}{p_{centre}(B_{\varphi},\lambda)}=\sqrt{1+(\tan{|\Delta\chi|)^2}}.
\label{eq33}
\end{equation}

\noindent For $\Delta\chi=60^{\circ}$ this ratio is equal to 2.

Clearly, it is not possible to explain all details of $p$ and $\chi$ in the our model of the sum of two spectral lines. But, above considerations tell us that the main characteristic features can be explained.

\subsection{The model of H$\alpha$ line as two emitting compact clouds located in the right and left part of the orbit}

The spectra of H$\alpha $ lines in the atlas of Smith et al.(2002) for many objects have sufficiently complicated structure -- the existence of separate peaks and asymmetric shapes. There are only a few objects with symmetric shapes. They have comparatively small widths as compared to other, more complicated spectra.

A line with a complicated shape frequently can be approximated as a sum of two or more lines with gaussian shapes. For this reason we present the model of two compact optically thick clouds located in the opposite parts of an orbit in the accretion disc (Fig.3). The advantage of this model compared with the previous one is that we can take into account all the depolarizing Faraday parameters $a, b_{\varphi} $ and $b_{\rho} $ in a simple analytical form.

Let us take the first cloud in the right part of the orbit, characterized by the azimuthal angle $\varphi$, and the second cloud in the left part, characterized by -$\varphi$. If it is necessary, locations of the emitting clouds can be chosen at arbitrary angles along the orbit. Our choice is the simplest for consideration. As usually, we write the Stokes parameters in the coordinate system with X-axis being perpendicular to the plane $({\bf nN})$, where the formulas have the simplest form. Further we will use the observed polarization degree $p$ and the total difference of the position angles between right and left parts of the spectral line $\Delta \chi $, which do not depend on the choice of the reference frame. We also include in our formulas the contribution of the continuum radiation ($I_c, Q_c, U_c$) in the region of spectral line. Then, the observed Stokes parameters are:

\begin{equation}
 I=I_c+I_{right}+I_{left},\,\,\,\,\xi=\frac{I_{right}}{I_{left}},
 \label{eq34}
\end{equation}

\[
 Q=Q_c+\frac{I_{right} p_{line}(\mu)(1-s\mu)g_{line}}{A_{+}}+
\]
\begin{equation}
+\frac{I_{left}p_{line}(\mu)(1-s\mu) g_{line}}{A_{-}},
\label{eq35}
\end{equation}

\[
 U=U_c+\frac{I_{right} p_{line}(\mu)(1-s\mu)(a+b_{\varphi})}{A_{+}}+
\]
\begin{equation}
+\frac{I_{left}p_{line}(\mu)(1-s\mu)(a-b_{\varphi})}{A_{-}},
\label{eq36}
\end{equation}

\[
A_{+}=g^2_{line}+(a+b_{\varphi})^2,
\]
\begin{equation}
A_{-}=g^2_{line}+(a-b_{\varphi})^2
\label{eq37}
\end{equation}

The explicit formulas for $Q_c$ and $U_c$ are as follows:

\[
Q_c=\frac{\sqrt{r+(g^2_c+b^2-a^2)}}{\sqrt{2}r}I_c p_c(\mu) (1-s\mu),
\]
\begin{equation}
U_c=\frac{\sqrt{r-(g^2_c+b^2-a^2)}}{\sqrt{2}r}I_c p_c(\mu) (1-s\mu).
\label{eq38}
\end{equation}

\noindent Here $r^2=(g^2_c+b^2-a^2)^2+4a^2g^2_c=g^4_c+2g^2_c(a^2+b^2)+(a^2-b^2)^2$. Introducing the notation $\eta_c=p_c(\mu)(1-s\mu)/p_c({\bf B},\mu)$, and using expressions (8) and (9), we can present formulas for $Q_c$ and $U_c$ in a simpler form:
\[
Q_c=\sqrt{1+\sqrt{1- \left(\frac{2ag_c}{\eta_c^2}\right)^2}}\frac{I_c p_c({\bf B},\mu)}{\sqrt{2}}
\to I_cp_c({\bf B},\mu),
\]

\begin{equation}
U_c=\sqrt{1-\sqrt{1- \left(\frac{2ag_c}{\eta_c^2}\right)^2}}\frac{I_c p_c({\bf B},\mu)}{\sqrt{2}}
\to 0.
\label{eq39}
\end{equation}

\noindent The last expressions are valid in the limit $(2ag_c)/\eta^2_c \to 0$.

In most sources from the catalog of Smith et al. (2002), the intensity $I_c$ is much smaller than $I_{right}+I_{left}$. For these cases one can neglect the contribution of the continuum radiation and the formulas for $p_{line}$ and $\chi_{line}$ acquire fairly simple form:

\begin{equation}
\tan{2\chi_{line}}=\frac{U_{line}}{Q_{line}}=\frac{a}{g_{line}}+\frac{b_{\varphi}}{g_{line}}\cdot\frac{1-\xi A_{+}/
A_{-}}{1+\xi A_{+}/A_{-}},
\label{eq40}
\end{equation}

\[
p_{line}=\frac{p_{line}(\mu)(1-s\mu)g_{line}(1+\xi A_{+}/A_{-})}{(1+\xi) A_{+}}\times
\]

\begin{equation}
\times \sqrt{1+(\tan{2\chi_{line}})^2}.
\label{eq41}
\end{equation}

In the right wing, where $\xi\simeq 0$, we have

\begin{equation}
 \tan{2\chi_{right}}=\frac{a+b_{\varphi}}{g_{line}}.
\label{eq42}
\end{equation}

\noindent For the left wing one finds the same expression with $(-b_{\varphi})$ instead of $b_{\varphi}$. Using formula (42), we can obtain expression for the difference of the position angles between the right and left wings of the line:

\[
\Delta\chi=\chi_{right}-\chi_{left}=
\]

\begin{equation}
=\frac{1}{2}\left(\arctan\frac{a+b_{\varphi}}{g_{line}}- \arctan\frac{a-b_{\varphi}}{g_{line}}\right).
\label{eq43}
\end{equation}

For the polarization degree in the right wing we derive the formula:

\begin{equation}
p_{right}=\frac{p_{line}(\mu)(1-s\mu)}{\sqrt{g_{line}^2+(a+b_{\varphi})^2}}.
\label{44}
\end{equation}

\noindent Analogously, for the left wing one replaces $b_{\varphi}$ with $(-b_{\varphi})$.

For the left wing the polarization degree is higher because the Faraday depolarization parameter $|a-b_{\varphi}|$ is lower than in the right wing. If $a=0$, the polarization degree $p_{right}=p_{left}$ and $\chi_{right}=-\chi_{left}$. Presence of the magnetic field $B_{\parallel}$ (parameter $a\neq 0$) diminishes the polarization degree both in the right and left wings of the line, and also diminishes the difference $|\Delta\chi|=|\chi_{right}-\chi_{left}|$. Besides, the functions $p_{line}(\lambda)$ and $\chi_{line}(\lambda)$ become asymmetric relative to the center of the line $\lambda_0$ (if the intensities of lines $I_{right}(\lambda)$ and $I_{left}(\lambda)$ are the same gaussian functions).

It is interesting to compare the polarization degrees in the wings and in the center of line. The general formula for ratio $p_{wing}/p_{center}$ is very complex and we consider only the case $a=0$, where this ratio reaches a maximum value. Taking $\xi=1$ for the center of the line and $\xi=0$ for the line wing, we obtain the following expression from the general formula (41):

\begin{equation}
\frac{p_{wing}}{p_{center}}=\sqrt{g_{line}^2+b_{\varphi}^2}.
\label{eq45}
\end{equation}

For $b_{\varphi}=5$ and $g_{line}=1$ this ratio is equal to 5.1, i.e. is considerably greater than the value from  formula (33). It is quite natural, because formula (33) describes the mean value of the effect. Clearly, the averaging procedure diminishes the effect. Physically this effect arises as a consequence of the Faraday rotation of the polarization plane. In the center of the line the rotations from the right and the left lines have opposite directions and the parameter $Q_{center}$ reaches a lower value than that in the line wing.

The most important conclusion from the theoretical consideration of the structure of broad lines in AGN is the treatment of the symmetry of the polarization degree $p_{line}(\lambda)$ as the result of the azimuthal magnetic field $B_{\varphi}$. If the symmetry of $p(\lambda)$ is considerably broken, one can consider that $B_{\varphi}\sim B_{\parallel}$, or the intensities of the right and the left emitting clouds are different.

\section{The magnetic field strength in a broad line region of AGN} 

The standard Unified Sheme for AGN includes a central source of continuum (accreting SMBH); a region close to the outer radius of the accretion disc emitting broad emission lines (broad line region -- BLR); a dusty rotating "torus" on parsec scales; and gas emitting narrow emission lines on a scale of tens to hundreds of parsecs, ionized through the open cone defined by the torus edge (Antonucci \& Miller 1985; Krolik \& Begelman 1988; Urry \& Padovani 1995).

The main unknown is the mechanism for the generation of the magnetic field during the process of accretion onto SMBH. Li (2002), Wang et al. (2002, 2003), Zhang et al. (2005) have studied the magnetic coupling process (MC) as an affective mechanism for transforming the kinetic energy of accreting gas into the magnetic energy. It is assumed that the disc is stable, perfectly conducting and Keplerian. The magnetic field on the black hole horizon is poloidal and varying as a power law with the distance from the central region.

Since the magnetic field on the horizon $B_H$ is brought and held by its surrounding magnetized matter of the accretion disc, there must exist the relation between the magnetic field strength near the BH horizon and the accretion rate $\dot{M}$.

As a result, the magnetic field strength on the event horizon $R_H=R_G(1+\sqrt{1-a_{*}^2})$ is determined by the relation between the magnetic energy and the accretion kinetic energy densities (see Li 2002; Wang et al. 2002):

\begin{equation}
 B_H=\frac{\sqrt{2k\dot{M}c}}{R_H}=\frac{\left(2kL_{bol}/\varepsilon c\right)^{1/2}}{R_G\left[1
+\sqrt{1-a_*^2}\right]}.
\label{eq46}
\end{equation}

\noindent Here $R_G=GM_{BH}/c^2$, the bolometric luminosity $L_{bol}=\varepsilon\dot{M}c^2$, $\dot{M}$ is the accretion rate, $c$ is the light velocity and $\varepsilon$ is the radiative efficiency calculated by numerical simulations of Novikov \& Thorne 1973, Krolik 2007, Shapiro 2007. The coefficient $k$ presents the inverse plasma parameter $k=P_{magn}/P_{gas}=1/\beta$, where $P_{gas}$ and $P_{magn}$ are the gas and the magnetic pressures, respectively. For the equipartition case $\beta=1$ and $k=1$.

Eq.(46) is easily transformed into:

\begin{equation}
 B_H=6.3\cdot 10^8\left(\frac{M_{\odot}}{M_{BH}}\right)^{1/2}\left(\frac{kl_E}{\varepsilon}\right)^{1/2}\frac{1}
{1+\sqrt{1-a_*^2}},
\label{eq47}
\end{equation}

\noindent where $l_E=L_{bol}/L_{Edd}$ and the Eddington luminosity $L_{Edd}=1.3\cdot 10^{38}\left(M_{BH}/M_\odot\right)$.

The basic problem is the relation between the magnetic fields strengths at the first stable circular orbit $R_{ms}$ and the event horizon $R_H$. The value of the radius $R_{ms}$ depends on the radius $R_G$ and the spin $a_*$, and can be presented in a form:

\begin{equation}
R_{ms}=q(a_*)R_G,
\label{eq48}
\end{equation}

\noindent where parameter $q>1$. For example, for a Schwarzschild black hole $q=6$ and for the Kerr BH with the spin $a_*=0.998$  $q=1.22$ (Murphy et al. 2009).

Reynolds, Garofalo \& Begelman (2006) argued that the plunging inflow can greatly enhance the trapping of large scale magnetic field on the black hole. Blandford (1990) has shown that the interaction of the large-scale magnetic field with the event horizon of rotating black hole can enhance the trapping of large-scale poloidal magnetic field on the horizon of the black hole, compared with the inner accretion flow and compared to the magnetic field strength derived from the relation between magnetic energy and accretion kinetic energy (Eq.(47)).

Recently Garofalo (2009) has showed that the dynamics of the plunge region of a thin black hole accretion disc and magnetic flux trapping can enhance the strength of the magnetic field threading the horizon by a significant factor. The results of his calculations were presented in fig.7 of Garofalo paper. It means that we obtain the following relation between the magnetic field strength at the first stable orbit $B_{ms}$ and the magnetic field strength at the event horizon of a black hole $B_H$:

\begin{equation}
 B_H=\eta(a_{*})B_{ms}.
\label{eq49}
\end{equation}

The coefficient $\eta$ can be obtained from fig.7 of the paper by Garofalo (2009). From this figure it follows that for $a_*=0.5$ the value is $\eta=5$ and for $a_*=0.0$ and $a_*=0.998$ we have $\eta=7.5$.

Numerical simulations have been used to study magnetic field generation in astrophysical objects. For example, the existence of large-scale dynamos in magneto-convection under the influence of shear and rotation has been studied by K\"{a}pyl\"{a}, Korpi \& Brandenburg (2008). These authors have shown that the saturation field strength reaches, practically, the equipartition level $B\approx0.7B_{eq}$, i.e. $k\approx0.5$. Taking into account the shear flows can increase the magnetic field  level. It means that the magnetization parameter can be equal to $k\approx1$.

We suggest that the magnetic field far inside in the accretion disc, and, especially, in the Broad Line Region (BLR) takes the toroidal form. Namely, the differential rotation in the accretion disc leads to an increase of the azimuthal field by winding up the poloidal field lines into the toroidal field lines (Bonanno \& Urpin 2007).

In astrophysical objects differential rotation is often associated with magnetic fields of various strength and geometry. If the poloidal field has a component parallel to the gradient of the angular velocity, then differential rotation can stretch toroidal field lines from the poloidal ones. In the presence of the magnetic field, differential rotation can be a reason for various MHD instabilities, especially if the field geometry is complex.

Mayer \& Pringle (2006) assumed that a dynamo process generates a local poloidal field $B_z$ in the accretion disc, and the magnitude of the poloidal component is small: $B_z \ll B_{\perp}$.

Usually one assumes that regular dependence of the magnetic field on the radius $R$ in the accretion disc exists from the first stable orbit $R_{ms}$, and that dependence has a power law form:

\begin{equation}
 B_{\perp}(R)=B_{ms}\left(\frac{R_{ms}}{R}\right)^n.
\label{eq50}
\end{equation}

We assume two values for the parameter $n$. The value $n=1$ derives the toroidal topology of the magnetic field (see Bonanno \& Urpin 2007). The value $n=5/4$ corresponds to the accretion process with hot accretion flows (Medvedev 2000).

For pure toroidal topology of the magnetic field in the accretion disc we have the depolarization parameter $a=0$, and in this case Eqs.(8) and (9) for non-turbulent case ($C=0$) are transformed into Eq.(14). Recall that, according to definitions (11), magnetic field $B(R_{BLR})$ can be determined, if the parameter $b$ is known from the observed polarization:

\begin{equation}
B(R_{BLR})=\frac{b}{0.8\lambda^2\sqrt{1-\mu^2}}.
\label{eq51}
\end{equation}

\noindent This field can be related to $B_{ms}$. In the case of a power-law dependence (50) with $n=1$, this relation becomes

\begin{equation}
 B(R_{BLR}) = B_{ms}\frac{R_{ms}}{R_{BLR}}=2.22\cdot 10^{-4}\left(\frac{M_9}{l_E}\right)^{1/2}q(a_{*})B_{ms}.
 \label{eq52}
\end{equation}

\noindent We used here Eq.(3) for determining $R_{BLR}$. Note that $M_9=M_{BH}/10^9M_{\odot}$. In this situation the depolarization parameter $b$ for H$\alpha$ wavelength ($\lambda=0.6563\mu$m) is:

\begin{equation}
 b=7.7\sqrt{1-\mu ^2}q(a_{*})\left(\frac{M_9}{l_E}\right)^{1/2}\left(\frac{B_{ms}}{10^5G}\right).
 \label{eq53}
\end{equation}

\noindent For hot accretion flows $n=5/4$ and the depolarization parameter $b$ becomes

\begin{equation}
 b=0.93\sqrt{1-\mu ^2} \left(\frac{M_9}{l_E}\right)^{5/8} q^{5/4}(a_*)\, \left(\frac{B_{ms}}{10^5G}\right),
 \label{eq54}
\end{equation}

\noindent where $q(a_{*})=R_{ms}/R_G$ and $R_G=GM_{BH}/c^2$. The explicit form of $q(a_{*})$ is given, for example, in Zhang et al. (2005).

Below we shall also consider the case of the equipartition between the gas and the magnetic pressures, i.e. $k \approx 1$. Namely, the magnetic coupling process corresponds to this case (Li 2002; Wang et al. 2003; Zhang et al. 2005; Ma, Wang \& Zuo 2006).

Using Eqs.(47) and (49), we transform relations (53) and (54) into the following forms:

\begin{equation}
 b = 1.53\sqrt{1-\mu ^2} \sqrt{\frac{k}{\varepsilon}}\, \frac{q(a_{*})}{\eta(a_{*})(1+\sqrt{1-a_{*}^2}\,)},
\label{eq55}
\end{equation}

\noindent and for $n=5/4$:

\[
 b = 0.19 \sqrt{1-\mu^2} \left(\frac{M_9}{l_E}\right)^{1/8} q^{1/4}(a_*) \sqrt{\frac{k}{\varepsilon}}\times
\]

\begin{equation}
 \times\frac{q(a_{*})}{\eta(a_{*})(1+\sqrt{1-a_{*}^2}\,)}.
 \label{eq56}
\end{equation}

Eqs.(55) and (56) allow us to estimate the radiation efficiency and therefore the rotation rate $a_*$ of an accreting black holes. Below we use the spectropolarimetric atlas of AGNs by Smith et al. (2002) for specific estimates.

\section{Magnetic field strength of Akn~120} 

According to Smith et al. (2002) the mean polarization degree in the continuum of Akn 120 is equal to $p_{c}\simeq 0.35\%$, and is equal to $\simeq 0.4\%$ in the H$\alpha$ line emission. We use here the data obtained by Smith et al. (2002) in 1998 October. The mean observed position angle have the same value for the continuum and line emission $\chi\simeq 76^{\circ}$. The inclination angle $i=48^{\circ}, \mu\simeq 0.67$. Its value has been derived by Braatz \& Gugliucci (2008) from water maser observations. The central black hole mass in Akn 120 is equal to $ M_{BH}\simeq 10^{7.74}M_{\odot}$ (see Peterson et al. 2004).

For the inclination angle $i=48^\circ$ the polarization in the continuum from the accretion disc without magnetic field is expected at the level $p_c(\mu)=1.26\%$ (Chandrasekhar 1960). This value is higher than the observed polarization degree and it means that the Faraday depolarization effect is really acting. The H$\alpha$ line spectrum shows the two-peak structure and can be represented as sum of two intensities with gaussian shapes. The difference of the positional angles $\Delta \chi$ is estimated to be between $70^{\circ}$ and $80^{\circ}$. The intensity of the continuum radiation reaches $\approx 18\%$ of the maximum line intensity near the centre. The behaviour of polarization in the continuum is very complex. It seems this behaviour occurs due to the existence of large-scale turbulent curls along the observed orbit. For that reason, it is more convenient to use the continuum - subtracted spectrum of Akn~120, presented in fig.24 of the mentioned Atlas by Smith et al. (2002).

We used formulas (31)--(37) to find the estimates for the parameters $a, b_{\varphi}, p_{line}(\mu)$ and $g_{line}$. Attempts to estimate these parameters under the assumption $p_c(\mu)\simeq p_{line}(\mu)$ were unsuccessful. We propose that in the line radiating clouds there is true absorption of radiation (the existence of dust particles). As it is known (see, for example, Silant'ev 1980), the existence of absorption gives rise to a considerable increase of polarization escaping from the optically thick atmosphere. This effect is the consequence of absorption creating a peak like form of escaping emission.

\begin{equation}
g_c\simeq 1,\,\,\,\, \frac{a}{g_{line}}\simeq 0,\,\,\,\, \frac{b_{\varphi}}{g_{line}}\simeq 3.55,\,\,\,\,
\frac{p_{line}(\mu)}{g_{line}}\simeq 4.07\%.
\label{eq57}
\end{equation}

\noindent Introducing first three parameters in Eq.(8), we obtain $g_{line}\simeq 0.975$. The estimate $g_{line}=0.975=1+C-s\mu$ gives the relation between $C$ and $s$. It means that the clouds are absorbing and the small scales are turbulent. The existence of large scale turbulence in clouds directly follows from very high line width. For the rest of radiation from the accretion disc outside the compact clouds we assume that absorption is absent and small scale turbulence is negligible ($C\approx 0$). From Eq.(57) it follows that $p_{line}\simeq 3.97\%$ and $b_{\varphi}\simeq 3.46$. The estimated value $p_{line}\simeq 3.97\%$ corresponds to the case when one takes into account the pure absorption of radiation by dust particles existing into emitting clouds. From Eq.(11) one can obtain an estimate of magnetic fields $B_{\parallel}\simeq 0$ G and $B_{\varphi}\simeq 14$ G.

\begin{figure}
 \includegraphics[width=84mm]{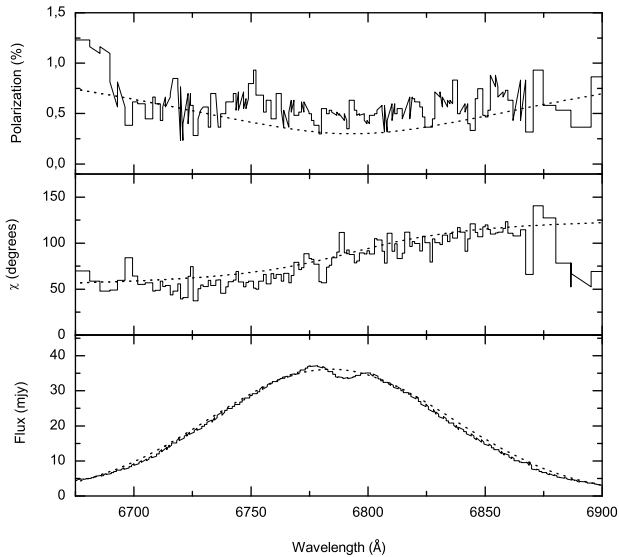}
 \caption{The comparison of observed (solid curves) and model data (dot curves) for Akn~120}
\end{figure}

In Fig.4 we present the observed intensity, polarization degree and variation of the azimuthal angle $\chi$, and our model results. It is seen that the model curves practically coincide with the observed values. The observed intensity is approximated as sum of two gaussian intensities from  symmetrically located clouds. The Doppler widths of these intensities are taken equal to $\Delta\lambda_T=$70{\AA}, the value $(u_k/u_{turb})\sin\varphi\sqrt{1-\mu^2}=0.3$, the ratio of maximum intensities of gaussians is $I_{left}/I_{right}=20/18 $. The centers of gaussian curves coincide with the observed places in Fig.24 of Smith et al. (2002).

The main problem is that the exact value of index $n$ is unknown and its value depends strongly on the model of the accretion disc. Pariev, Blackman \& Boldyrev (2003) suggest the following interval of values of this index $1\leq n\leq2$. The value $n=1$ corresponds to toroidal magnetic field (see Bonanno \& Urpin 2007).

To estimate the magnetic field strength at the last inner circular orbit $B_{ms}$ we need to know the rotation rate of the central supermassive black hole (parameter $a_*$). For a Schwarzschild black hole with $a_*=0$ and $q=6$ and for the toroidal magnetic field ($n=1$) we obtain from Eq.(49) $B_{ms}=14.5\cdot10^3G$. Now we can estimate the magnetic field strength $B_H$ at the horizon of the central black hole using the results of calculations by Garofalo (2009). He has calculated the ratio of the horizon-threading magnetic field and the magnetic field in the accretion disc as a function of the black hole spin. According to this calculations, for $a_*=0$ the ratio $B_H/B_{ms}=7.5$ and $B_H=10.7\cdot10^4G$.

For the spin value of $a_*=0.5$, $\varepsilon=0.081$, $q=4.25$ (Novikov \& Thorne 1973) we obtain $B_{ms}=6.8\cdot10^3G$ and $B_H=3.4\cdot10^4G$. At last, for $a_*=0.998$, $q=1.22$, $\varepsilon=0.32$ we have $B_{ms}=2.8\cdot10^4G$ and $B_H=14\cdot10^4G$.

Our results demonstrate that for a given value of the polarization degree the magnetic field strength at the inner radius $r_{ms}$ (and therefore on the horizon radius) is stronger for a Kerr black hole compared to Schwarzschild one. This result means also that for black holes with the same magnetic field strengths the degree of polarization for Kerr black holes is larger than for Schwarzschild black holes (see Silant'ev et al. 2011).

\section{Magnetic field strength of Mrk~6} 

According to Ho, Darling \& Greene  (2008), the mass of the central black hole in Mrk 6 is $\log(M_{BH}/M_\odot)=7.97\pm0.5$,  the ratio of the bolometric luminosity to the Eddington value is $\log(L_{bol}/L_Edd)=-1.72$. The inclination angle  is $i=62^\circ.7, \mu=0.46$ (Ho et al. 2008). It means that the standard (Sobolev--Chandrasekhar) magnitude of  the polarization degree is $p_c(\mu)=2.52\%$. The observed mean polarization has been found at the level of  $p_c=0.90\pm0.03$, $p(H_\alpha)=0.85\pm0.04$ (Smith et al. 2002). The most remarkable fact is the jump  of the mean position angle for two observational seasons of Feb 97 and Oct 98: $\Delta\chi\approx 25^\circ$. It is interesting that the value of the mean polarization remained the same. This jump occurred over 1-2 years, which is too short a time for such a large object as the accretion disc near the supermassive black hole. Thus, this problem remains unsolved.

Let us return now to the analysis of the data in the H$\alpha$ line. First of all, we see that the polarization degrees in the right and the left sides of the spectrum are practically equal - the left side has $p_{line}\simeq 1.5\%$, and in the right side $p_{line}\simeq 1.4\%$. From the theoretical considerations in section 2, it is clear that such a symmetrical form of the polarization degree can occur if the magnetic field $B_{\parallel}$ is much less than the azimuthal magnetic field $B_{\varphi}$. A small decrease of polarization in the right-hand part is due to the reason that small value of the $B_{\parallel}$ component, directed to an observer, coincides with direction of $B_{\varphi}$.   In this situation the polarization of radiation slightly decreases compared to the left-hand part of the orbit, where the aforementioned magnetic fields are directed opposite to each other. Considering polarization near the line center, we assume $p_{line}(centre)\simeq 0.5\%$. We also assume that in the center of the line the intensity from the left part of the orbit is approximately equal to that from the right part. The value of the intensity of the continuum radiation in Mrk 6 is relatively small, and we neglect it in our computation. Using formulas (34)--(36), one obtain:

\begin{equation}
g_c\simeq 1, \,\,\,\, \frac{a}{g_{line}}\simeq 0.1,\,\,\,\, \frac{b_{\varphi}}{g_{line}}\simeq 2.615,\,\,\,\,
\frac{p_{line}(\mu)}{g_{line}}\simeq 3.9\%.
\label{eq58}
\end{equation}

\noindent Substitution of these parameters in Eq.(8) gives $g_{line}\simeq 1.0006$, i.e. practically 1. As in the case of Akn~120, the polarization $p_{line}\simeq 3.9\%$ implies that in H$\alpha$-clouds there exists the absorption of radiation. Under the assumption of dipole radiation we have $q_c\simeq 0.01$ and the parameter $s\simeq 0.17$. Because $g_{line}=1.0006\simeq 1=1+C-0.17\cdot 0.46$ we find that the small scale turbulent parameter $C\simeq 0.08$. For the difference $\Delta\chi $ of the position angles between the left and the right wings of the spectral line the expression (43) gives $\Delta \chi = 42^{\circ}$. This value is consistent with the observational results. Eqs.(11) and (58) give estimate: $B_{\parallel}\simeq 0.6$G and $B_{\varphi}\simeq 8.5$G. Note that expression (58) supposes that $I_cp_c\ll I_{right, left}\,p_{line}(\mu)$.

Now let us estimate the magnetic field strength in the accretion disc of Mrk 6 using the results of the polarimetric observations from Smith et al. (2002). Eqs. (53)--(56) allow us to derive the magnetic field strength $B_{ms}$ at the inner radius of the accretion disc: $B_{ms}=(1.72\times10^4/q)G$.  For a Swarzschild black hole, when $q=6$, the magnetic field $B_{ms}=2.9\cdot10^3G$. For a Kerr black hole  with $a_*=0.998$ the parameter $q=1.22$ and the magnetic field strength $B_{ms}=1.4\cdot10^4G$.  According to Garofalo (2009, fig.7) the magnetic field strength at the horizon of the supermassive black hole is $B_H=1.4\cdot10^4G$ for $a_*=0$ and $B_H=10^5G$ for $a_*=0.998$.

\section{Magnetic fields of Mrk~985 and IZw1}

The intensity spectrum of Mrk 985 has a two-peak shape and the spectrum of the polarization degree $p_{line}(\lambda)$ is quite symmetric (the $p_{line}(left)\simeq 1.27\%$ and $p_{line}(right)\simeq 1.16\%$). In the centre of line $p_{line}(centre)\simeq 0.5\%$. The mean polarization of the continuum radiation is 1.12\%. Using formulas (33)--(37), we find that

\[g_c\simeq 1, \,\,\,\, \frac{a}{g_{line}}\simeq 0.114,
\]
\begin{equation}
 \frac{b_{\varphi}}{g_{line}}\simeq 2.035,
\,\,\, \frac{p_{line}(\mu)}{g_{line}}\simeq 2.75\%.
\label{eq59}
\end{equation}

\noindent We see that a nearly symmetric form of $p_{line}(\lambda)$ implies that $a\ll b_{\varphi}$, i.e. $B_{\parallel}\ll B_{\varphi}$.

In the atlas of Smith et al. (2002) there is no information on the inclination angle $i$. It is interesting to estimate this angle assuming that $g_{line}\simeq 1$. Substituting parameters $g_c=1, a=0.114$ and $b_{\varphi}=2.035$ into formula (8) gives the value $p_c(\mu)=2.54\%$. This implies an estimate $i\simeq 64^{\circ}, \mu\simeq 0.44$. Using the value $p_c(\mu)=2.54\%$ we can calculate the value $g_{line}$, corresponding to this polarization. This calculation demonstrates that the parameter $g_{line}$ acquires the value $g_{line}\simeq 1.0013$. The value $p_{line}(\mu)=2.75$ at $\mu=0.44$ takes place at $q_a\simeq 0.01$ ($s=0.17$). The value $g_{line}\simeq 1.0013=1+C-0.17\cdot 0.44$ occurs at $C\simeq 0.08$. Using Eq.(11) and the value $\mu\simeq 0.44$, we find the estimates: $B_{\parallel}\simeq 0.75$ G and $B_{\varphi}\simeq 6.6$ G.

From Fig.15 in the atlas of Smith et al. (2002) we see that the difference of position angles between the right wing and the centre of line is equal $\Delta\chi\simeq 31-33^{\circ}$. From general theory one finds that $\tan{2\Delta\chi}=a/g_{line}+b_{\varphi}/g_{line}$. Our estimates (59) give this value for the angle difference. The estimates of magnetic fields at distances $R_{ms}$ and $R_H$ can be obtained analogously as in the previous sections.

Now let consider the AGN IZw1. This object has $p_c\simeq 0.67\%$, $p_{line}(left)\simeq 0.7\%$, $p_{line}(centre)\simeq 0.2\%$ and $p_{line}(right)\simeq 0.9\%$. The form of the polarization spectrum is slightly more asymmetric than in Mrk~985. Using general formulas (33)--(35) we find the following estimates:

\[
 g_c\simeq 1, \,\,\, \frac{a}{g_{line}}\simeq 0.52,
\]
\begin{equation}
 \frac{b_{\varphi}}{g_{line}}\simeq 3.905,
\,\,\,\frac{p_{line}(\mu)}{g_{line}}\simeq 3.18\%.
\label{eq60}
\end{equation}

As in the case of Mrk~985, we found, that $g_{line}\simeq 1$ and $p_c(\mu)\simeq 2.68\%$. This corresponds to $\mu\simeq 0.44$. The value $p_{line}\simeq 3.9\%$ occurs at the absorption degree $q_a\simeq 0.01, s\simeq 0.17$. As a result, the small scale turbulence parameter $C\simeq 0.08$.

As in the previous cases, using Eq.(11) and the value $\mu=0.44$ we can estimate the magnetic field strength for IZw 1: $B_{\parallel}=3.42$G and $B_{\perp}=12.7$G. It should be noted that in these cases we estimated the inclination angle $i$ from the analysis of the polarization data.

The obtained estimates of the magnetic field strengths in BLR of these AGNs are presented in Table 1. The last value in the table is close to the one estimated by Afanasiev et al.(2011) where the polarimetric observations were made only for the continuum emission and did not include the emission from the broad line region.

\begin{table}
 \centering
 \caption{The obtained estimates of the magnetic field strengths.}
 \begin{tabular}{lrr}
 \hline
 AGN     & $B_{\parallel}$[G] & $B_{\perp} = B_{\varphi}$ [G] \\
 \hline
 Akn 120 & 0                  & 14 \\
 Mrk 6   & 0.6                & 8.5 \\
 Mrk 985 & 0.75               & 6.6 \\
 IZw 1   & 3.42               & 12.7 \\
 \hline
 \end{tabular}
\end{table}

\section{Conclusions}

For many objects of Smith et al. (2002) spectropolarimetric atlas the polarization spectra of a broad H$\alpha$-line $p_{line}(\lambda)$ have a characteristic minimum at the center of the line and different maxima at the left and right wings. Usually the wing polarizations are higher than those in the nearby continuum. For many objects the position angle changes continuously from the left wing to the right one. We develop a new theoretical explanation for these features, different from the original explanation of Smith et al. (2002, 2004, 2005), taking into account that accretion discs can be magnetized.

Smith et al. explain the characteristic features of the H$\alpha$-line assuming that the observed polarization is due to single scattering of non-polarized radiation from the BLR on two types of scattering clouds - a polar cloud around the radio jet, and clouds in the equatorial region. It appears (see Smith et al. 2005) that their mechanism has two characteristic features: a relatively low amplitude ($|\Delta\chi|\le 20-40^{\circ}$) of the position angle rotation from one line wing to the other one, and the need for a very high electron temperature ($T\sim 10^6$ K)\ in a nearby scattering cloud. In their atlas there are cases with both low $\Delta\chi < 20^{\circ}$ and high $\Delta\chi\sim 80^{\circ}$ position angle rotations. The need for the high electron temperature arises from the observed polarization minimum in the line core. Besides, they neglect the intrinsic linear polarization of the radiation in BLR.

In our mechanism both effects can be explain by a single cause - the Faraday rotation of the polarization plane. The wide line width results from turbulence, which is related to the Keplerian rotation in the orbit. Clearly both explanations take place in reality.

The basis of our explanation is the assumption that both the continuum radiation and the spectral line emission originate in an optically thick magnetized accretion disc around the center of an AGN. The observed characteristic shape of the line polarization appears as a result of the Faraday rotation of the polarization plane in the accretion disc having a normal magnetic field $B_{\parallel}$ and an azimuthal field $B_{\varphi}$.

We propose that the regions of the line emission can be represented as a comparatively dense absorbing turbulent clouds rotating with the Keplerian velocity around the center of the AGN. These clouds are flattened, optically thick and magnetized. They emit the polarized radiation in accordance with the Milne problem law. The observed emission line is a sum of radiation from clouds rotating in the right and the left sides of the orbit. Due to Doppler displacements, emission from the one side is, as a whole, reddened ($\lambda \ge \lambda_0$), and emission from the other side has the opposite $\lambda$-displacement. The Faraday rotations by azimuthal magnetic field $B_{\varphi}$ in the left and the right sides of the orbit are opposite and, as a result, in the center of line the sum of emissions is less polarized than in the wings. The continuous rotation of the position angle $\chi$ from one wing of line to the opposite wing arises for the same reason. The projection of the normal magnetic field $B_{\parallel}$ along the line of sight gives an additional Faraday rotation. It is the same in both sides of the radiating orbit. This additional rotation in one side of orbit increases the total Faraday rotation, and in opposite side decreases the total rotation. For this reason $B_{\parallel}$ magnetic field gives rise to an asymmetric (relative to the central wavelength $\lambda_0$) profiles for both the polarization degree $p_{line}(\lambda)$ and the position angle $\chi_{line}(\lambda)$. The presented theory allows us to estimate the components $B_{\varphi}$ and $B_{\parallel}$ in the broad line emission regions, and also in nearby regions of the continuum radiation.

If the polarizations in the left and the right wings are slightly different, then the value  $B_{\parallel}\ll B_{\varphi}$. This helps us to estimate the $B_{\varphi}$-component from more simple formulas for continuum polarization. Objects in which this is the case are ESO 141-635, IZw1, Mrk 6, Mrk 290, Mrk 985 and NGC 5548. The objects with a strong asymmetry of $p_{line}(\lambda)$ are characterized by magnetic fields $B_{\varphi}\sim B_{\parallel}$. Such objects are Akn 120, Akn 564,KUV 18217+6419, Mrk 304, Mrk 335, Mrk 841, MS1849.2-7832 and NGC 4593. Other objects have fairly complex polarization spectra, they appear to be distorted by large-scale turbulent motions.

Using the estimated values of the magnetic field in broad line regions {\bf (usually $B_{\parallel}\ll B_{\varphi}$ and $B_{\varphi}\sim 10$ G)}, one can estimate the magnetic field $B_{ms}$ at the last stable orbit near the black hole, and then the field $B_H$ at the radius of the event horizon. These estimates are dependent on the different assumptions about the slope of the power-law distribution of the magnetic field inside the accretion disc. We have used the most common assumptions to obtain the values of the magnetic field $\sim 10^4-10^5$ G. These values of the magnetic field are in a good agreement with other estimates. Thus, we determined the magnetic field strengths in various places in the accretion discs of AGNs from the real observational polarization data.

\section*{Acknowledgments}

This research was supported by the program of Prezidium of RAS No.21, the program of the Department of Physical Sciences of RAS No.16, by the Federal Target Program "Scientific and scientific-pedagogical personnel of innovative Russia" 2009--2013 and the Grant from President of the Russian Federation ''The Basic Scientific Schools'' NSh-1625.2012.2.

\end{document}